\documentclass[11pt]{article}

\usepackage[margin=1in]{geometry}
\usepackage{booktabs}
\usepackage{graphicx}
\usepackage{amsmath}
\usepackage{natbib}
\usepackage{hyperref}

\title{Implied ETF Carry Rates and the Limits of Arbitrage in Segmented Bitcoin Markets}
\author{Mindy L. Mallory\\Associate Professor, Purdue University\\\href{mailto:mlmallor@purdue.edu}{mlmallor@purdue.edu}}
\date{May 2026}

\newcommand{\NObs}{386}
\newcommand{\MeanWedge}{2.58}
\newcommand{\WedgeFive}{-4.77}
\newcommand{\MedianWedge}{2.52}
\newcommand{\WedgeNinetyFive}{10.42}

\begin{document}

\maketitle

\begin{abstract}
This paper estimates the carry embedded in listed IBIT options and compares it with the carry embedded in matched CME bitcoin futures. Put-call parity recovers an implied forward on the ETF; BlackRock's daily holdings file maps each ETF share into bitcoin units; and CME futures prices and BRRNY, a U.S.-close bitcoin reference rate, provide the corresponding futures-market carry. The difference in the carry implied by these two products likely reflects friction resulting from inability to cross margin spot bitcoin and CME futures, resulting in higher and more variable carry in CME bitcoin futures than the carry implied by IBIT listed options on IBIT. In the current selected-strike IBIT sample of \texttt{\NObs} date-bucket observations, the mean wedge is \texttt{\MeanWedge}\% and the median wedge is \texttt{\MedianWedge}\%, both measured in annual percentage points. The result is consistent with segmented collateral and margin systems limiting arbitrage between regulated bitcoin-exposure venues. 
\end{abstract}

\noindent\textbf{Keywords:} Bitcoin ETFs; put-call parity; futures basis; limits to arbitrage; cross-margining

\noindent\textbf{JEL codes:} G12; G13; G18

\section{Introduction}

Spot bitcoin ETFs created a regulated equity-market instrument for holding bitcoin exposure. Listed options on those ETFs created a second instrument: a way to infer the forward price embedded in the ETF-options rail. CME bitcoin futures provide a third price for related bitcoin exposure. These instruments reference the same economic asset, but they do not sit inside one integrated collateral and margin system. If arbitrage were frictionless, their implied forward prices should align after accounting for fees, maturities, and benchmark timing. Persistent differences are therefore informative about the cost of moving risk across market infrastructures. A cash-and-carry trade in bitcoin requires spot bitcoin financing and custody on one side and futures margin on the other. An ETF-options implementation can keep the ETF and option legs inside the securities and listed-options infrastructure, although the exact margin offset depends on account type, broker, and clearing rules.

This paper measures that difference for BlackRock's iShares Bitcoin Trust (IBIT). Put-call parity gives an ETF-implied forward price from matched call and put prices. The daily IBIT holdings file gives the number of bitcoin represented by each ETF share.\footnote{The IBIT holdings file is available from BlackRock's iShares holdings endpoint: \url{https://www.ishares.com/us/products/333011/fund/1467271812596.ajax?dataType=fund&fileName=IBIT_holdings&fileType=csv}.} Dividing the ETF-implied forward by that holdings ratio gives an ETF-options-implied bitcoin forward. I compare the resulting carry with the carry on matched CME bitcoin futures, using BRRNY as the U.S.-close bitcoin spot benchmark for the daily CME carry leg and as the benchmark reference for IBIT.\footnote{BRR and BRRNY are published by CF Benchmarks at \url{https://www.cfbenchmarks.com/data/indices/BRR} and \url{https://www.cfbenchmarks.com/data/indices/BRRNY}.} The wedge is CME futures carry minus fee-adjusted ETF-options carry. A positive value means that CME futures embed more annualized carry than the ETF-options rail.

Prior work uses put-call parity deviations to study financing frictions in equity markets \citep{ofek2003limited,cremers2010deviations}. Recent work shows that crypto carry can be large when arbitrage capital is segmented and margin requirements bind \citep{schmeling2023cryptocarry,siriwardane2022segmented}. This paper connects those ideas by using listed spot-bitcoin-ETF options to estimate a regulated-market bitcoin financing wedge. The object is useful because it is observable from listed option prices and fund holdings, and because it speaks directly to the funding and margin channels emphasized in the crypto-carry literature.

The estimates also have a market-design interpretation. Mature derivatives markets have explicit mechanisms for recognizing offsetting risk across some related positions. OCC describes cross-margining as a way to recognize intermarket hedged positions cleared by different organizations, reducing initial margin requirements and settlement liquidity needs for eligible participants \citep{occ_cross_margin}. The OCC--CME program applies this principle to eligible cleared options and futures positions \citep{cme_occ_2009}. The bitcoin result suggests more explicit recognition of hedged bitcoin exposures across the ETF-options and futures rails could lower intermediation costs. 

\section{Cash-and-Carry Mechanism}

A bitcoin cash-and-carry trade buys spot bitcoin and sells a bitcoin futures contract. Ignoring transaction costs and margin frictions, the trader finances the spot position, holds or custodies the bitcoin, and receives the futures premium through the short futures position. CME bitcoin futures are cash settled rather than physically delivered. For standard CME bitcoin futures, the final settlement reference rate is BRR, the 4 p.m. London CME CF Bitcoin Reference Rate.\footnote{CME states that standard Bitcoin and Micro Bitcoin futures settle to BRR, while Bitcoin Friday futures settle to BRRNY. CME also states that daily settlement prices for larger and Micro cryptocurrency futures are based on a VWAP of Globex trades between 2:59 p.m. and 3:00 p.m. Central Time. See \url{https://www.cmegroup.com/articles/faqs/frequently-asked-questions-cryptocurrency-futures.html}.} That final-settlement rule is distinct from the daily carry measurement in this paper. The empirical series uses daily futures closes, IBIT closes, and option quotes observed on U.S. trading dates. The CME carry leg is therefore measured as the annualized premium of the matched futures price over BRRNY, the U.S.-close bitcoin reference rate.

The ETF-options version uses the same economic logic but different market plumbing. The trader can hold IBIT shares and sell an IBIT synthetic forward by selling a call and buying a put with the same strike and expiration. Put-call parity gives the forward price embedded in that option pair. Because IBIT represents a fractional bitcoin position, the fund holdings file maps the ETF forward into bitcoin units. The ETF-options carry is the annualized premium of this implied bitcoin forward over the IBIT share price, adjusted for the fund expense ratio. This trade does not require custody of spot bitcoin, but it still depends on ETF financing, option-market liquidity, margin requirements, and the ability of the broker and clearing system to recognize offsetting risk.

The comparison is useful because the two implementations transfer similar bitcoin price risk through different clearing and collateral systems. The CME implementation uses futures margin. The ETF-options implementation uses securities and listed-options infrastructure and is tied to IBIT, whose benchmark index is BRRNY.\footnote{BlackRock reports the IBIT benchmark index as the CME CF Bitcoin Reference Rate - New York Variant, Bloomberg ticker BRRNY. See \url{https://www.ishares.com/us/products/333011/ishares-bitcoin-trust}.} The wedge measures how much carry differs across those two implementations after converting the ETF leg into bitcoin units.

\section{Data and Measurement}

The analysis combines IBIT option quotes and underlying prices from OptionMetrics IvyDB US, IBIT holdings files from BlackRock, SOFR from the Federal Reserve Bank of New York, CME bitcoin futures histories from Barchart, and BRRNY as the U.S.-close bitcoin spot benchmark. The analysis focuses on IBIT because it is the dominant spot bitcoin ETF and has the cleanest option-market depth in the current sample.

IBIT uses BRRNY as its benchmark index. Standard CME bitcoin futures final-settle to BRR rather than BRRNY, while Bitcoin Friday futures settle to BRRNY. Nevertheless, I use BRRNY because it aligns the bitcoin benchmark with the daily closing prices of the CME bitcoin futures. The BRR and BRRNY are based on the same bitcoin spot prices, they just provide the same bitcoin spot price index as a daily series based on different points in time. The daily carry measure is built from U.S.-date futures closes, ETF closes, and option quotes. BRR is the correct benchmark for final settlement of standard CME bitcoin futures, but it is not time-aligned with the daily U.S.-close prices used here.

For each date $t$, expiration $T$, and strike $K$, I merge calls and puts on common strike and compute the put-call-parity forward
\[
F^{ETF}_t(K,T)=K+e^{r_t\tau_{t,T}}\left(C_t(K,T)-P_t(K,T)\right),
\]
where $C_t(K,T)$ and $P_t(K,T)$ are option midquotes and $\tau_{t,T}$ is calendar time to expiration divided by 365. IBIT options are American style, so early exercise is a potential concern. The current design addresses this conservatively through sample selection rather than a structural American-option adjustment. The sample keeps contracts with 14 to 90 calendar days to expiration, absolute moneyness below 5\%, bid-ask spread below 10\% of the option midquote, and open interest of at least 100 contracts on both option legs.

After filtering, the selected-strike series keeps one tradable pair per date, expiration, and maturity bucket. The selected strike is the surviving pair closest to the money, with ties broken by higher pair open interest and then lower spread. The selected expiration is the one closest to the bucket target: 22 calendar days for the 14--30 day bucket and 45 calendar days for the 31--60 day bucket. This avoids averaging across many strikes while retaining a stable daily measurement object. Since we will compare the IBIT-implied carry with the CME-implied carry we do not want to artificially smooth the IBIT carry series by averaging implied carry across strikes. 

Let $q_t$ be bitcoin per IBIT share, computed from BlackRock's reported bitcoin holdings divided by shares outstanding. The ETF-options-implied bitcoin forward is $F^{ETF}_t(K,T)/q_t$. The ETF-options carry is the exact effective annualized premium of the selected ETF forward over the IBIT close, plus the annual fund expense ratio, which is 0.25\%.\footnote{BlackRock reports the current IBIT expense ratio as 0.25\% on the iShares IBIT product page: \url{https://www.ishares.com/us/products/333011/ishares-bitcoin-trust}.}

The CME carry is the effective annualized premium of the matched CME futures close over BRRNY, annualized over the matched futures contract's own remaining duration. The wedge is
\[
w_t(T)=y^{CME}_t(T)-y^{ETF,adj}_t(T).
\]
All reported carries and wedges are annual percentage points.

\section{Results}

Table~\ref{tab:tab:summary_statistics} reports the main summary statistics. The usable selected-strike sample contains \texttt{\NObs} date-bucket observations. The mean wedge is \texttt{\MeanWedge}\%, the median is \texttt{\MedianWedge}\%, the 5th percentile is \texttt{\WedgeFive}\%, and the 95th percentile is \texttt{\WedgeNinetyFive}\%. These magnitudes are economically meaningful. They are not small pricing residuals around zero.

\begin{table}

\caption{\label{tab:tab:summary_statistics}Summary statistics for the selected-strike IBIT wedge series.}
\centering
\begin{tabular}[t]{rrrrrr}
\toprule
observations & Mean (\%) & SD (\%) & P05 (\%) & Median (\%) & P95 (\%)\\
\midrule
386 & 2.581 & 4.716 & -4.767 & 2.521 & 10.418\\
\bottomrule
\end{tabular}

\begin{minipage}{0.94\linewidth}
\footnotesize\emph{Notes:} The sample keeps one near-the-money call-put pair and one representative option expiration in each of the 14--30 day and 31--60 day maturity buckets. The wedge is measured in annual percentage points as the exact effective annualized CME carry minus the exact effective annualized fee-adjusted IBIT-implied carry. The CME carry is annualized over the matched futures contract's own duration.
\end{minipage}
\end{table}

\begin{table}

\caption{\label{tab:tab:wedge_by_maturity}IBIT wedge summary by option-maturity bucket.}
\centering
\begin{tabular}[t]{lrrrr}
\toprule
maturity\_bucket & Observations & Mean (\%) & Median (\%) & SD (\%)\\
\midrule
14-30d & 193 & 2.222 & 2.061 & 5.527\\
31-60d & 193 & 2.939 & 2.669 & 3.713\\
\bottomrule
\end{tabular}

\begin{minipage}{0.94\linewidth}
\footnotesize\emph{Notes:} Each date-bucket observation uses a single selected expiration and a single near-the-money strike. Entries are annual percentage points, not log units. The 61--90 day bucket is excluded because the long end of the current IBIT sample is thin and produces unstable exploratory comparisons.
\end{minipage}
\end{table}

\begin{figure}[htbp]
\centering
\includegraphics[width=0.9\textwidth]{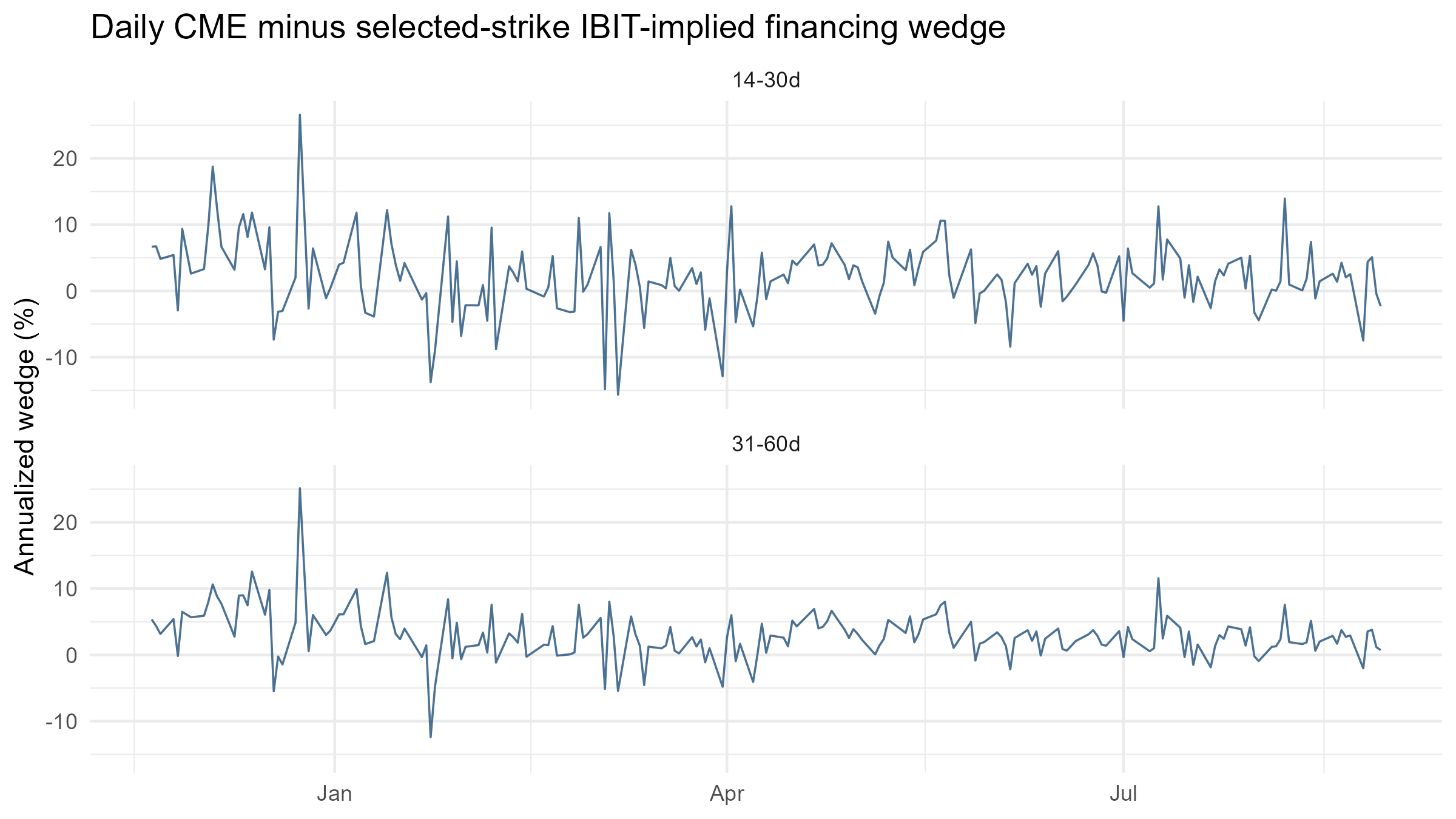}
\caption{Daily IBIT wedge series.}
\label{fig:wedge_timeseries}
\par\smallskip
{\footnotesize \emph{Notes:} The wedge is matched CME bitcoin futures carry minus fee-adjusted IBIT-implied carry. It is reported in annual percentage points. Positive values indicate that the CME rail embeds more carry than the ETF-options rail.}
\end{figure}

Figure~\ref{fig:wedge_timeseries} plots the wedge by maturity bucket, and Table~\ref{tab:tab:wedge_by_maturity} reports the corresponding bucket-level summary statistics. Positive values mean that CME futures price bitcoin exposure more richly than the fee-adjusted ETF-options rail.

\begin{figure}[htbp]
\centering
\includegraphics[width=0.9\textwidth]{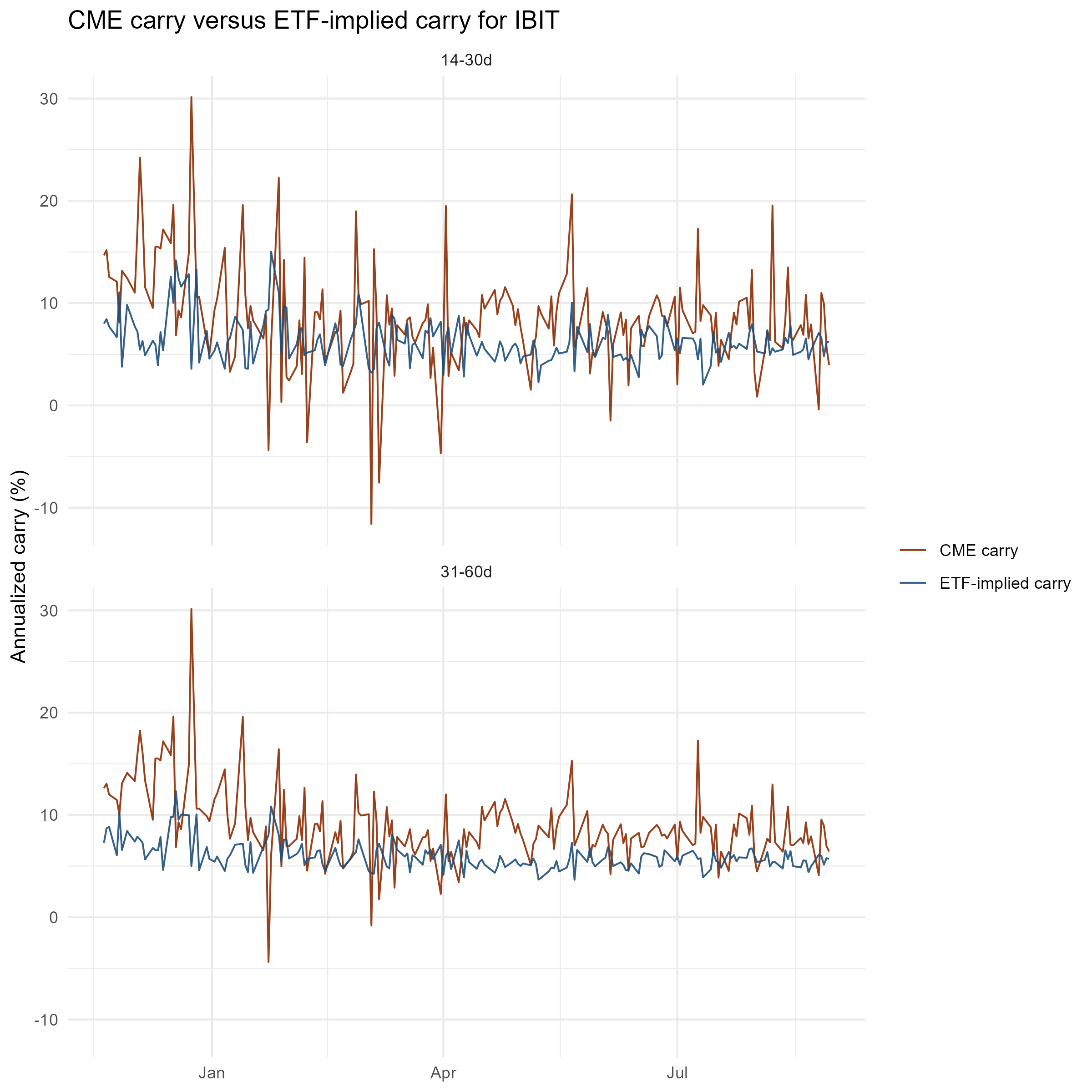}
\caption{CME carry and fee-adjusted IBIT-implied carry by maturity bucket.}
\label{fig:carry_comparison}
\par\smallskip
{\footnotesize \emph{Notes:} Each line is a daily annualized carry. CME carry is computed from the matched futures contract and BRRNY. ETF carry is computed from the selected put-call-parity-implied IBIT forward, scaled by bitcoin per share and adjusted for the IBIT expense ratio.}
\end{figure}

Figure~\ref{fig:carry_comparison} plots the two carry legs directly. The comparison shows whether the wedge is coming from movements in CME carry, movements in ETF-implied carry, or both. A persistent gap is consistent with segmentation between the futures rail and the ETF-options rail. It is also consistent with margin and collateral systems not treating hedged bitcoin exposures as a single integrated position.

\section{Conclusion}

Listed spot-bitcoin-ETF options provide a direct way to estimate the carry embedded in the ETF-options rail. Comparing that carry with matched CME bitcoin futures carry gives a simple measure of cross-rail segmentation in regulated bitcoin markets. In the current IBIT sample, the average CME-minus-IBIT carry difference is \texttt{\MeanWedge}\% per year. The result is consistent with segmented collateral and margin systems limiting arbitrage. It also motivates consideration of whether greater recognition of offsetting bitcoin exposures could improve capital efficiency in the carry trade.

\section*{Declaration of Interest}

The author declares no known competing financial interests or personal relationships that could have appeared to influence the work reported in this paper.

\section*{Funding}

This research received no specific grant from any funding agency in the public, commercial, or not-for-profit sectors.

\section*{Data and Code Availability}

The data used in this study are not publicly available because the option and futures data include licensed vendor data. Code used to produce the analysis can be made available by the author upon reasonable request, subject to vendor data restrictions.

\section*{Declaration of Generative AI and AI-Assisted Technologies in the Writing Process}

During the preparation of this work, the author used OpenAI tools to assist with manuscript organization, language editing, code review, and repository audit. The author reviewed and edited the content as needed and takes full responsibility for the content of the publication. All empirical results reported in the manuscript are generated by the repository code and stored data files.

\bibliographystyle{apalike}
\bibliography{references_frl}

\end{document}